\documentclass[10pt,conference]{IEEEtran}

\usepackage{lipsum}
\usepackage{cite}
\usepackage{amsmath,amssymb,amsfonts}
\usepackage{balance}
\usepackage{algorithm}
\usepackage{subcaption}
\usepackage{graphicx}
\usepackage{tabularx}
\usepackage{array}
\usepackage{relsize}
\usepackage{textcomp}
\usepackage{xcolor}
\usepackage{braket} 
\usepackage{algpseudocode} 

\newcommand{\framework}{\textsc{DART-Q}}

\usepackage{graphicx}
\usepackage{textcomp}
\usepackage{xcolor}
\def\BibTeX{{\rm B\kern-.05em{\sc i\kern-.025em b}\kern-.08em
    T\kern-.1667em\lower.7ex\hbox{E}\kern-.125emX}}

\begin{document}

\title{\framework\ : A Deadline-Driven Framework for Real-Time QLDPC Decoding\\}


\author{
\IEEEauthorblockN{
Ameya S. Bhave\IEEEauthorrefmark{1},
Navnil Choudhury\IEEEauthorrefmark{2},
Kanad Basu\IEEEauthorrefmark{2}
}
\IEEEauthorblockA{
\IEEEauthorrefmark{1}Department of Electrical and Computer Engineering,
The University of Texas at Dallas, Richardson, TX, USA\\
\IEEEauthorrefmark{2}Department of Electrical, Computer, and Systems Engineering,
Rensselaer Polytechnic Institute, Troy, NY, USA\\
Email: \IEEEauthorrefmark{1}asb240006@utdallas.edu,
\IEEEauthorrefmark{2}choudn3@rpi.edu,
\IEEEauthorrefmark{2}basuk@rpi.edu
}
}

\maketitle

\begin{abstract}
Real-time quantum error correction places the classical decoder inside the fault-tolerant control loop under strict timing and memory constraints. For quantum low-density parity-check (QLDPC) codes, practical deployment therefore depends not only on correction performance, but also on timely decoding under deadlines, finite on-chip memory, and time-varying load. However, existing decoder studies primarily emphasize correction performance without exposing operational viability under these constraints. We present \framework{}, a real-time QLDPC decoding framework that treats windowed workloads as discrete arrival, queueing, service, and completion events. \framework{} models each decode request as a deadline-driven online service job with queueing and non-preemptive Earliest Deadline First scheduling. It supports configurable admission control, service times, and bounded rescue policies. Through controlled studies of the SRAM-fit transition, tail latency, overload, and a capacity-scaling extension, \framework{} isolates the effects of memory pressure, rescue selectivity, admission control, and pooled service capacity on timely decoding. Our results show that real-time decoder viability is governed by state organization, overload policy, and service capacity. A cached-summary state organization lowers the SRAM-fit boundary by \(4\times\) relative to an edge-centric baseline. Under overload, relaxing the backlog cap increases queued work by approximately \(20.1\times\) and worsens p99 latency by approximately \(17.6\times\), with little gain in useful throughput. In contrast, doubling decoder capacity reduces the MissRate from \(97.64\%\) to \(0.98\%\) and improves p99 latency from \(3.861\,\mathrm{ms}\) to \(100\,\mu\mathrm{s}\). These results position \framework{} as a framework for exposing the regime changes that determine real-time QLDPC decoder viability under deadlines, finite memory, and time-varying load.
\end{abstract}


\section{Introduction}
\label{sec:introduction}

\begin{figure}[!t]
  \centering
  \includegraphics[width=\linewidth]{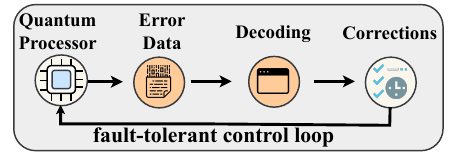}
  \caption{Real-time decoding in the fault-tolerant control loop.}
  \label{fig:intro_control_loop}
  \vspace{-5mm}
\end{figure}

Fault-tolerant quantum computing depends on quantum error correction (QEC) to preserve logical information over long computations \cite{terhal_2015}. In this setting, decoding lies in the live control loop, as shown in Figure~\ref{fig:intro_control_loop}. Detector events or syndrome data produced by the quantum processor arrive continuously, and the classical decoder must return corrections within tight time budgets \cite{Skoric_2023,caldwell2025platformarchitecturetightcoupling}. This requirement is especially important for quantum low-density parity-check (QLDPC) codes, which have emerged as a promising path to lower logical-qubit overhead \cite{breuckmann_eberhardt_2021,Bravyi_2024}. Their practical viability depends not only on correction performance, but also on decoding behavior under deadlines, finite memory, and time-varying load \cite{gong2024lowlatencyiterativedecodingqldpc,maurer2025realtimedecodinggrosscode,maurya2026fpgatailoredalgorithmsrealtimedecoding}. 

Most decoder studies focus on correction performance, asymptotic complexity, or average runtime \cite{breuckmann_eberhardt_2021,roffe_2020,delfosse2021unionfinddecoderquantumldpc,Grospellier_2021,leverrier2022decodingquantumtannercodes}. These metrics remain important, but they do not fully capture service-level failure modes in real-time settings, where hard instances can create long tails, backlog growth, and deadline misses that render decoder outputs unavailable at the required control boundary \cite{Skoric_2023,caldwell2025platformarchitecturetightcoupling}. Recent studies have examined decoder-specific hardware feasibility, latency-tail conditions, and low-latency decoder-feedback paths for particular implementations \cite{gong2024lowlatencyiterativedecodingqldpc,maurer2025realtimedecodinggrosscode,maurya2026fpgatailoredalgorithmsrealtimedecoding,bascones2025exploring,liu2026scalableopensourceqec}. In this work, we conduct a complementary system-level study of memory pressure, queueing, admission control, rescue semantics, and service capacity to assess timely decoding as a service.

We present \framework{}, a discrete-event framework for comparative analysis of real-time QLDPC decoding under deadlines, finite on-chip memory, and time-varying load. It adopts a deadline-driven service view in which decode instances become time-constrained service requests and queueing, scheduling, and admission control determine whether corrective updates are delivered on time. This view makes deadline misses, drops, queue growth, and tail latency explicit measures of decoder viability. Rather than serving as a cycle-accurate platform model, \framework{} isolates the operating transitions and policy tradeoffs that shape timely decoding. We use it to study the SRAM-fit transition, workload-sensitive tail latency, overload under admission control, and a capacity-scaling extension.

The main contributions of this paper are as follows:
\begin{itemize}
    \item \textbf{Systems framework.}
    We present \framework{}, a discrete-event framework for deadline-driven comparative analysis of real-time QLDPC decoding under finite memory and time-varying load.
 
    \item \textbf{Service abstraction.}
    We cast windowed QLDPC decoding as a deadline-driven online service with arrivals, deadlines, queueing, admission control, and non-preemptive EDF scheduling.

    \item \textbf{Deadline-driven regime analysis.}
    Across SRAM-fit-transition, tail-latency, overload, and extended capacity-scaling regimes, we evaluate how state organization, rescue selectivity, admission control, and pooled capacity shape real-time decoder viability.

    \item \textbf{Decoder viability insights.}
    We show that decoder viability is governed not only by correction performance or average runtime, but also by state organization, rescue selectivity, admission control, and pooled service capacity, supporting a service view of QLDPC decoding.

\end{itemize}


\section{Background and Motivation}
\label{sec:background_motivation}


\subsection{Decoding in the Control Loop}
\label{sec:bm_control_loop}

Fault-tolerant quantum execution produces a continuous stream of syndrome or detector events from repeated stabilizer measurements. A classical decoder consumes this stream and returns a corrective action, typically as a recovery operation or Pauli-frame update. This step lies on the runtime feedback path between measurement and continued logical execution \cite{terhal_2015,Skoric_2023}. Because measurements continue while the computation evolves, the decoder must keep up with the incoming stream. It must therefore do more than infer a valid correction: it must return that correction promptly to keep the corrective state aligned with the ongoing computation. Once decoding enters this feedback path, low latency becomes an operational requirement alongside correction quality \cite{Skoric_2023,caldwell2025platformarchitecturetightcoupling,battistel2023realtimedecoding}.

\subsection{Timing Variability, Tails, and Backlog}
\label{sec:bm_tails_backlog}

Timely decoding depends on more than matching the long-run arrival rate. A real-time decoder must complete each arriving decode instance within the time budget of its control epoch. When service times vary across instances, average throughput alone does not prevent deadline failure. This variability can arise from both algorithmic and architectural causes. Decode effort may change with syndrome weight, locality, or iteration count, while contention and memory stalls can further extend service time. Under streaming arrivals, even a small fraction of slow windows can build a queue. That backlog then increases waiting time for later windows, so deadline misses can occur even when average utilization appears acceptable. Real-time feasibility therefore depends on tail latency, queue growth, and overload behavior, not only on average throughput \cite{Skoric_2023,caldwell2025platformarchitecturetightcoupling}.

\subsection{Memory Footprint and SRAM Fit Boundary}
\label{sec:bm_sram_fit_boundary}

A key source of timing variability is the decoder's working-set size. Message-passing decoders such as Belief Propagation (BP) maintain persistent state across iterations, including intermediate messages and auxiliary data \cite{roffe_2020,muller2025improvedbeliefpropagationsufficient}. For BP-style QLDPC decoders, decoder organization therefore directly affects memory footprint and on-chip fit. Finite on-chip memory is also a practical constraint in real-time decoder implementations \cite{maurya2026fpgatailoredalgorithmsrealtimedecoding,maurer2025realtimedecodinggrosscode}. When the working state exceeds the available on-chip memory budget, accesses extend to lower-tier memory, increasing traffic and service latency. We refer to the resulting operating limit as the SRAM fit boundary.

This transition has direct timing consequences. Off-chip accesses increase latency, raise bandwidth pressure, and can widen the service-time tail. The effect is not only a gradual slowdown. Once the working set exceeds effective on-chip capacity, service cost can shift materially because traffic rises and timing becomes less predictable. The SRAM fit boundary therefore links decoder organization to real-time behavior through persistent state footprint and memory traffic, rather than through correction performance alone.


\subsection{Toward Decoding as a Service}
\label{sec:bm_service_bridge}

Taken together, these observations motivate a service-oriented view of real-time decoding. In the fault-tolerant control loop, the central challenge is to produce the right correction on time under continuous arrivals, variable service times, and memory-sensitive execution. We therefore model real-time decoding as a deadline-driven online service. Decode instances become jobs arriving over time, each with a service requirement and a timing constraint. Queueing, scheduling, and overload behavior then determine whether corrective updates are delivered in time. The next section defines this abstraction more rigorously.


\section{Problem Formulation}
\label{sec:problem_formulation}



\label{subsec:pf_service_model}
We formalize real-time decoding of QLDPC codes as a deadline-driven online service, as shown in Figure~\ref{fig:pf_timeline_queue}. Block~1 represents the time-ordered syndrome or detector-event stream, Block~2 partitions that stream into fixed-duration windows of length $W$ to form primary decode jobs $J_i$, and Block~3 represents the decoding service under deadlines. Each job has an arrival time $a_i$, a service requirement, and an absolute deadline $d_i$. Jobs may wait in a queue before service, and scheduling and admission policies determine service order and overload behavior \cite{Skoric_2023,caldwell2025platformarchitecturetightcoupling}. This intentionally minimal, system-facing abstraction retains the timing, queueing, and control features needed to study decoder viability while remaining agnostic to any specific decoder pipeline or hardware implementation. The job definition and timing semantics, queueing model, and service outcomes are detailed in the subsections that follow.



\subsection{Job Definition and Timing Semantics}
\label{subsec:pf_job_timing}
\begin{figure}[!t]
  \centering
  \includegraphics[width=\linewidth]{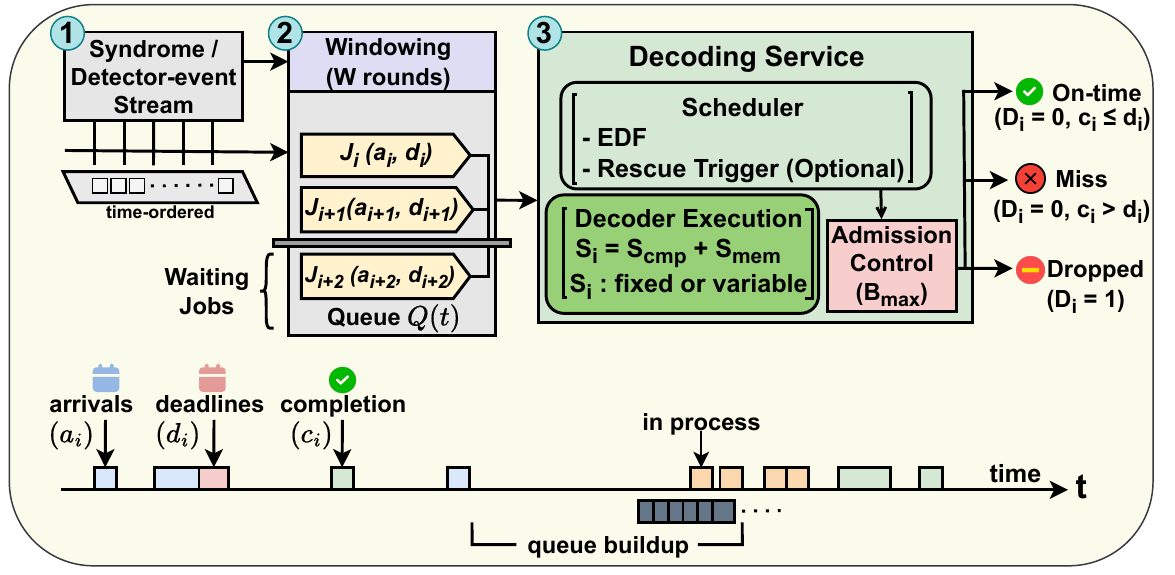}
  \caption{Real-time decoding as a service. A streaming syndrome or detector-event trace is partitioned into decode jobs, queued, and served under deadlines to produce corrective updates.}
  \label{fig:pf_timeline_queue}
  \vspace{-2mm}
\end{figure}

We index primary decode jobs by $i \in \{1,2,\dots n \}$ in window-arrival order. This order is defined because the primary workload is formed by non-overlapping fixed-duration windows over a single time-ordered syndrome stream. A primary job $J_i$ is formed in the windowing stage of Block~2 from a window of $W$ extraction rounds from the syndrome (or detector-event) workload, i.e., one decode request for the data accumulated over that window, as shown in Figure~\ref{fig:pf_timeline_queue}. The job input $x_i$ is the syndrome or detector-event data observed over that window, and the output $y_i$ is the corrective update produced by the decoder and returned to the control stack, typically a Pauli-frame update over the $n$ data qubits. Job $J_i$ arrives at time $a_i$ and is assigned an absolute deadline $d_i$ as:
\begin{equation}
d_i \triangleq a_i + \Delta
\label{eq:pf_absolute_deadline}
\end{equation}
where $\Delta > 0$ is a predefined slack budget. The job requires a service time $S_i$ and if admitted, it completes at time $c_i$. For admitted jobs, we define deadline offset $(L_i)$ and tardiness $(T_i)$ as:
\begin{equation}
L_i \triangleq c_i - d_i, \qquad
T_i \triangleq \max(0, L_i).
\label{eq:pf_lateness_tardiness}
\end{equation}
A deadline miss occurs when an admitted job completes after its deadline, i.e., when $L_i > 0$. In this abstraction, a late decode output is treated as unavailable at the required control boundary and therefore as a missed service outcome rather than a usable corrective action. For dropped jobs, $(c_i, L_i, T_i)$ are undefined. We model service time $S_i$ as:
\begin{equation}
S_i \approx S_{\mathrm{cmp}}(x_i) + S_{\mathrm{mem}}(x_i;\mathcal{S}),
\label{eq:pf_service_decomp}
\end{equation}
where $S_{\mathrm{cmp}}(x_i)$ is the workload-driven compute effort and $S_{\mathrm{mem}}(x_i;\mathcal{S})$ is the memory and traffic effects induced by the decoder working set $\mathcal{S}$. Variability in $S_i$ reflects event density, bounded-iteration behavior, and memory or bandwidth stalls.

\subsection{Queueing and Scheduling Model}
\label{subsec:pf_queue_scheduling}
We model the decoder as a single logical server with a waiting queue to capture the dominant shared bottleneck governing service progress, such as memory bandwidth beyond the on-chip budget or a pipeline-limited stage \cite{maurer2025realtimedecodinggrosscode,maurya2026fpgatailoredalgorithmsrealtimedecoding,caldwell2025platformarchitecturetightcoupling}. Let $Q(t)$ denote the queued backlog at time $t$, that is, the number of admitted jobs waiting for service. Figure~\ref{fig:pf_timeline_queue} highlights backlog through \(Q(t)\) and the queue-buildup timeline, while Block~3 shows the queue, scheduler, and admission-control path that governs service progression. Larger \(Q(t)\) increases waiting time and can push completion past the absolute deadline even when average utilization appears acceptable.

When the queue is nonempty, a scheduling policy $\pi$ determines which admitted job is served next. As a canonical baseline for deadline-driven workloads, we use non-preemptive earliest-deadline-first (EDF). EDF selects the queued job with the smallest deadline $d_i$ \cite{liu1973scheduling}. To limit queued backlog under overload, we allow admission control through a cap $B_{\max}$ applied to queue length just before arrival. On arrival, primary job $J_i$ is dropped if:
\begin{equation}
D_i =
\begin{cases}
1, & \text{if } Q(a_i^-) \ge B_{\max},\\
0, & \text{otherwise},
\end{cases}
\label{eq:pf_backlog_cap}
\end{equation}
where $D_i = 1$ indicates a dropped primary job and $Q(a_i^-)$ is the queued backlog immediately before arrival of $J_i$.

The base formulation is defined on primary jobs derived from syndrome windows. Some decoder designs may also insert bounded auxiliary jobs into the same queue, $e.g.$, to perform limited follow-up work triggered by the primary decode path. When present, auxiliary jobs have their own arrival, service, and deadline attributes and are scheduled under the same policy $\pi$. They add queueing delay and resource contention, with service-level metrics normalized over primary job arrivals unless stated otherwise.

\subsection{Service Objectives and Outcome Metrics}
\label{subsec:pf_objectives_metrics}

Our objective is to keep deadline misses and drops rare, tail response time under control, and backlog bounded over the primary decode stream. We distinguish drops from misses because a drop is an explicit overload-control action taken at arrival, whereas a miss is a late completion of an admitted primary job. Auxiliary work items, when present, affect the primary-stream metrics indirectly through the load they add unless reported separately. Figure~\ref{fig:pf_timeline_queue} summarizes the three service outcomes: on-time completion, miss, and drop. We define the miss indicator $(M_i)$ for primary job $J_i$ as:
\begin{equation}
M_i \triangleq \mathbb{I}[D_i = 0]\cdot \mathbb{I}[L_i > 0],
\label{eq:pf_miss_indicator}
\end{equation}
and, over $N$ primary job arrivals, we report the rates of missed and dropped jobs as:
\begin{equation}
\mathrm{MissRate}(N) \triangleq \frac{1}{N}\sum_{i=1}^N M_i, \quad
\mathrm{DropRate}(N) \triangleq \frac{1}{N}\sum_{i=1}^N D_i.
\label{eq:pf_miss_drop_rates}
\end{equation}

We also report Goodput, the fraction of arrivals that are admitted and complete on time, together with response time for admitted primary jobs as:
\begin{equation}
\mathrm{Goodput}(N) \triangleq \frac{1}{N}\sum_{i=1}^{N}
\mathbb{I}[D_i=0]\cdot \mathbb{I}[c_i \le d_i], 
\ 
R_i \triangleq c_i - a_i.
\label{eq:pf_goodput_response_time}
\end{equation}

We summarize latency using response-time quantiles (\textit{e.g.}, $p99$), together with tardiness summaries derived from Eq.~(\ref{eq:pf_lateness_tardiness}). We also report backlog statistics derived from $Q(t)$, including maximum observed backlog and threshold-based backlog-duration summaries~\cite{openai_chatgpt_2026}.
\section{\framework{} System Design}
\label{sec:system_design}

\begin{figure*}[t]
  \centering
  \includegraphics[width=0.9\linewidth]{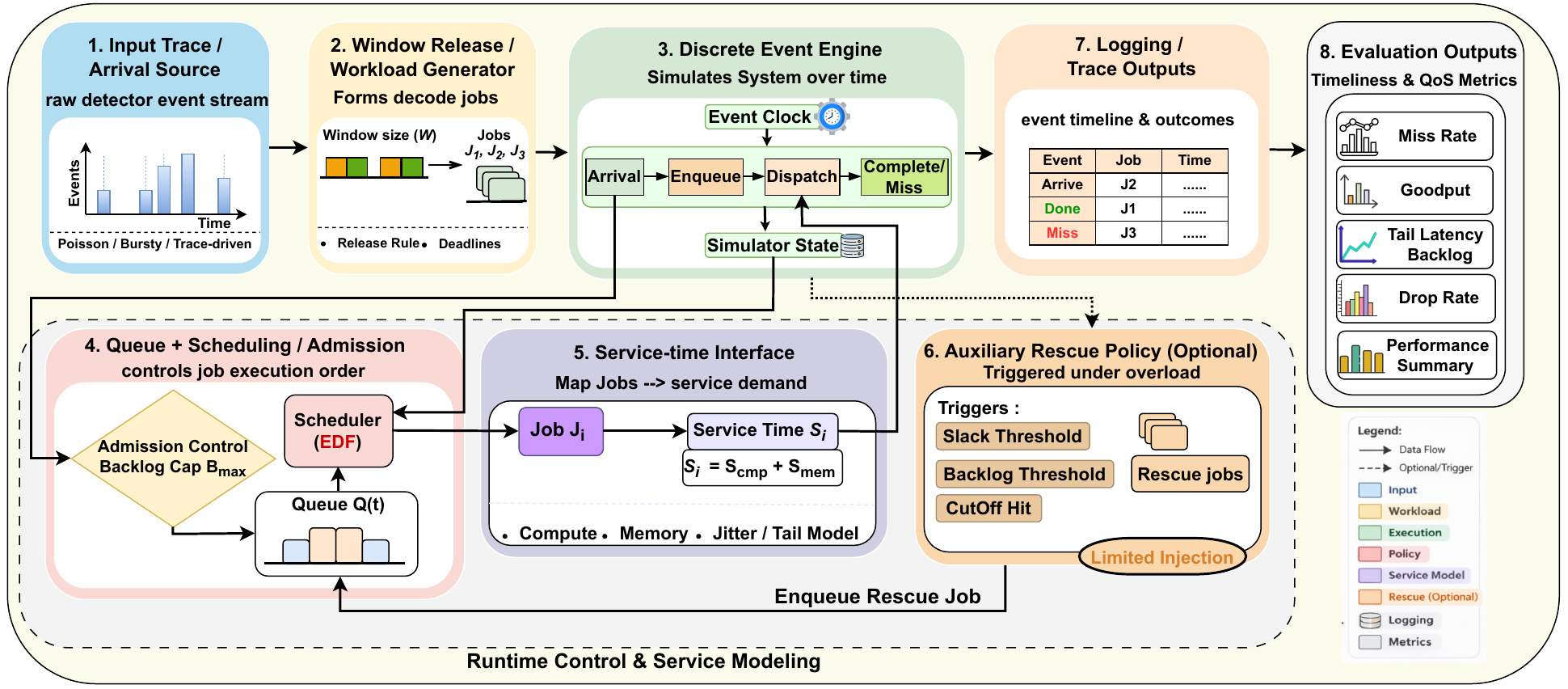}
  \caption{\framework{} architecture and measurement flow. Raw detector-event traces are released as $W$-round decode jobs and executed by a discrete-event engine with queueing, scheduling, and backlog-cap admission control. A service-time interface assigns compute and memory demand, while optional bounded rescue injects extra work under overload. Per-job outcomes and queue traces are recorded to derive timeliness, backlog, and QoS metrics.}
  \label{fig:sd_runtime_architecture}
    \vspace{-2mm}
\end{figure*}

\label{sec:sd_overview_mapping}


This section instantiates the deadline-driven decoding service of Sec.~\ref{sec:problem_formulation}. Figure~\ref{fig:sd_runtime_architecture} summarizes the runtime architecture across Blocks 1–8. The subsections that follow describe the execution model, service-time mapping, scheduling with overload control, and logging support used to recover the service-level metrics of Sec.~\ref{subsec:pf_objectives_metrics} and the cost-model quantities of Sec.~\ref{sec:cost_models_metrics}~\cite{openai_chatgpt_2026}.
\subsection{Execution Model}
\label{sec:sd_sim_core}


\framework{} implements the job and timing semantics of Sec.~\ref{subsec:pf_job_timing} and the queueing model of Sec.~\ref{subsec:pf_queue_scheduling} using a discrete-event execution engine. In Figure~\ref{fig:sd_runtime_architecture}, Blocks~1--2 provide the input trace and window-release logic, Block~3 advances the simulation through arrival, enqueue, dispatch, and completion events, and Block~4 realizes admission control, the explicit waiting queue, and scheduling. \framework{} maintains a global simulation clock and processes a priority-ordered event list. A primary decode job $J_i$ is instantiated from the workload representation of a fixed-duration window together with timing attributes $(a_i,d_i)$. Although the base formulation assumes a windowed decode stream, \framework{} can also emulate stressed or aggregated conditions by offsetting release times, releasing multiple windows together, or merging pre-windowed input streams. When an admitted primary job completes, \framework{} records its completion time $c_i$ and outcome. Response time, deadline offset, and tardiness then follow as described in Sec.~\ref{subsec:pf_job_timing}.

Service proceeds through a single logical server with an explicit waiting queue. The single-logical-server model captures the dominant shared bottleneck, such as memory bandwidth beyond the on-chip budget or a pipeline-limited stage. In Figure~\ref{fig:sd_runtime_architecture}, the explicit waiting queue and scheduler appear in Block~4, while dispatch and completion events are realized through the discrete-event engine in Block~3. The framework absorbs internal hardware parallelism into the effective service time $S_i$ rather than modeling it as separate servers. Whenever the server becomes available, the active non-preemptive policy $\pi$ selects the next admitted job from the queue. The framework also records backlog evolution over time so that queueing and backlog-derived statistics can be recovered consistently with Sec.~\ref{subsec:pf_queue_scheduling}. The current framework uses a logical single-server queue as a conservative baseline when latency is governed primarily by a dominant shared resource constraint.

\subsection{Service-Time Model}
\label{sec:sd_service_model}

For each admitted primary job, \framework{} assigns service time $S_i$ through a configurable mapping from job attributes and hardware parameters to service demand. Depending on the selected fixed- or variable-service model of Sec.~\ref{sec:cost_models_metrics}, this mapping assigns either fixed or job-dependent values. In Figure~\ref{fig:sd_runtime_architecture}, it appears in Block~5 as the service-time interface that maps a scheduled job to service demand. The mapping preserves the queueing abstraction of Sec.~\ref{subsec:pf_queue_scheduling} and follows the decomposition of Eq.~(\ref{eq:pf_service_decomp}) by separating workload-driven compute effort from memory and traffic effects induced by decoder working set $\mathcal{S}$. The mapping is parameterized by hardware-relevant quantities, including working-set size, effective on-chip SRAM budget, and bandwidth or stall parameters that determine the memory-related component of service time. The same interface also captures workload-dependent compute variation derived from job payload statistics. This design lets the framework represent both stable service regimes and regimes with heavier service-time tails under the same queueing abstraction. Further details are provided in Sec.~\ref{sec:cost_models_metrics}.

\subsection{Scheduling and Overload Control}
\label{sec:sd_policies}

\framework\ implements the scheduling and admission abstraction of Sec.~\ref{subsec:pf_queue_scheduling} through a configurable policy $\pi$. In Figure~\ref{fig:sd_runtime_architecture}, this corresponds to the queue, scheduler, and admission path in Block 4, together with the optional rescue path in Block 6. The default baseline is the non-preemptive EDF policy of Sec.~\ref{subsec:pf_queue_scheduling}; specific alternative priority and rescue policies are introduced with the corresponding evaluation regimes in Sec.~\ref{sec:evaluation}. Admission control implements the backlog-cap rule of Eq.~(\ref{eq:pf_backlog_cap}). On arrival, a primary decode job is dropped when $Q(a_i^-)\ge B_{\max}$, and the framework records the drop indicator $D_i=1$ immediately. The backlog-cap rule limits queued backlog under overload and makes the tradeoff between deadline misses and drops explicit.

Beyond the base formulation of primary decode jobs, \framework{} also introduces auxiliary work items to study limited recovery mechanisms. In Figure~\ref{fig:sd_runtime_architecture}, auxiliary work items appear as the rescue path in Block 6. A rescue attempt is modeled as an auxiliary item with its own arrival, deadline, and service attributes. \framework{} injects rescue work back into the same queueing path and schedules it under the active policy alongside primary jobs, so that any benefit from limited recovery is evaluated against the additional queueing delay and contention it introduces. In each run, rescue injection is triggered by a policy condition, such as a slack threshold, a queue-occupancy threshold, or a bounded budget on additional service time. The queue-occupancy threshold is measured in jobs and the additional-service budget limits the auxiliary recovery effort. Section~\ref{sec:evaluation} evaluates the impact of auxiliary work items on the primary-stream metrics defined in Sec.~\ref{subsec:pf_objectives_metrics}.
\subsection{Logging, Measurement \& Traceability}
\label{sec:sd_artifacts}

During evaluation, each run of \framework\ produces a fully traceable record of simulated service behavior. In Figure~\ref{fig:sd_runtime_architecture}, Block~7 records event traces and Block~8 aggregates them into service-level and QoS metrics. Each run records: (i) the full run configuration, including workload, policy, and service-model parameters, (ii) per-item timing outcomes \((a_i,d_i,S_i,c_i)\) together with admission and completion outcomes, and (iii) backlog evolution over time together with summary utilization statistics. When the selected service model exposes decomposed timing, the logs also retain compute, memory, and total service components.

These records allow the service-level metrics of Sec.~\ref{subsec:pf_objectives_metrics} to be recomputed exactly from logged traces. MissRate and DropRate are computed over primary job arrivals. Response-time quantiles are computed over admitted primary jobs that complete service. Backlog statistics are derived from the logged $Q(t)$ trajectory. When auxiliary work items are present, the framework logs them explicitly so that their overhead can be separated from the primary job stream. All stochastic choices are seeded and recorded, so reported results are reproduced by deterministic post-processing of the archived run configuration, per-job traces, and logged queue trajectory rather than by manual run selection. As an extension of this baseline, the same timing abstraction can also be instantiated with multiple identical decoder service instances behind a shared queue to study capacity scaling without changing the core timing model.


\section{Cost Models and Metrics}
\label{sec:cost_models_metrics}


We instantiate the service-time interface of Sec.~\ref{sec:sd_service_model} with cost models for memory traffic, workload-sensitive compute effort, and their composition. Figure~\ref{fig:cm_hw_arch} summarizes the system studied in this section and links decoder state organization to memory footprint, the SRAM fit boundary, service time, and queueing behavior. The models are intentionally comparative rather than platform-predictive: they isolate how decoder organization, on-chip state budget, bandwidth, and workload variability shape service time under shared assumptions. Through the queueing model of Sec.~\ref{sec:problem_formulation}, these effects propagate to deadline misses, drops, and latency tails. The framework supports both fixed-service and variable-service models, including broader-tailed regimes.

\subsection{Traffic Model}
\label{sec:cm_traffic_v1}
\begin{figure}[!t]
  \centering
  \includegraphics[width=\linewidth]{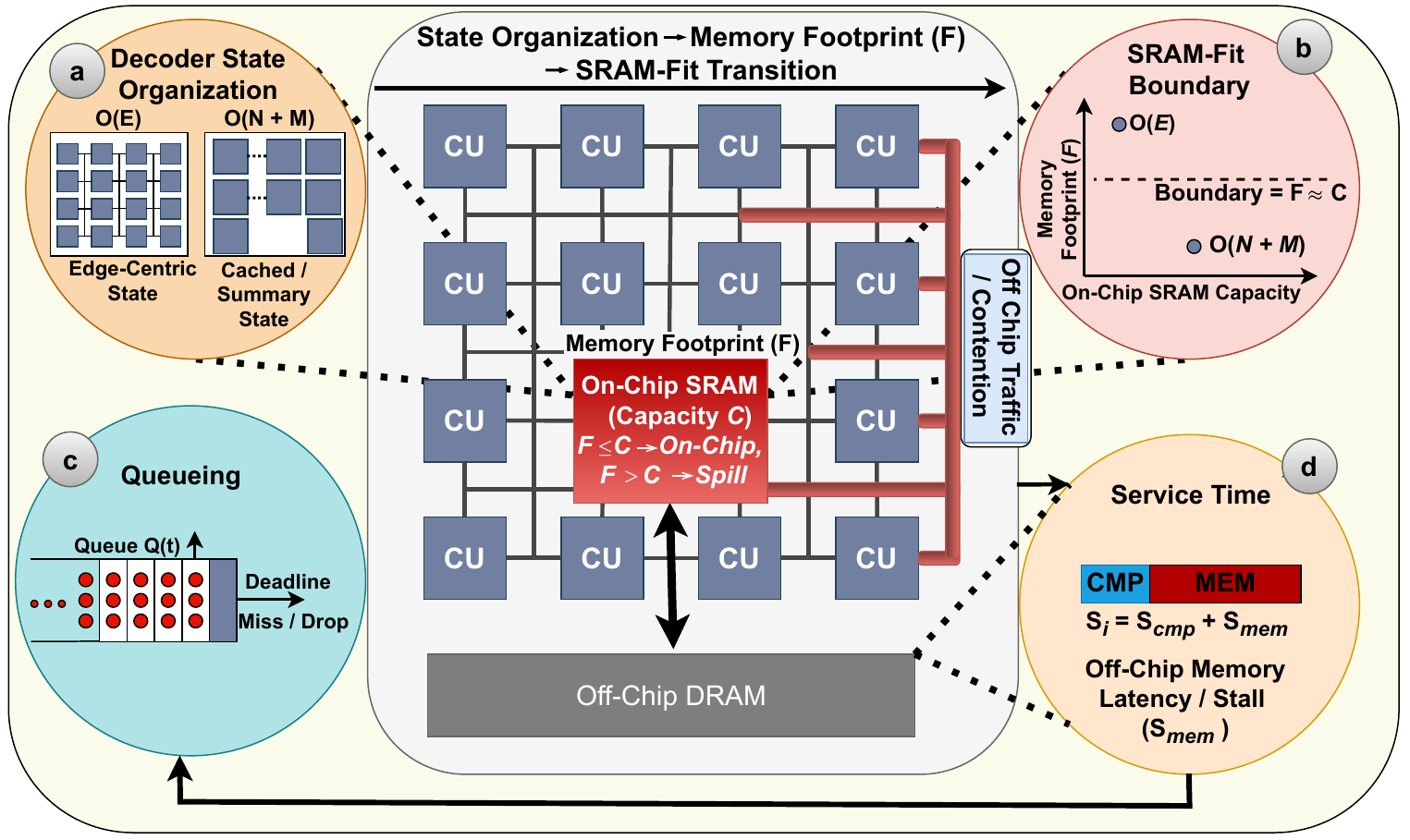}
  \caption{Traffic model and SRAM-capacity knee. (a) Decoder state organization sets the persistent memory footprint. (b) The knee occurs when the aligned footprint exceeds the on-chip SRAM budget. (c) Spill-induced off-chip traffic slows service and drives queue growth, misses, or drops. (d) Service time is the compute floor plus a memory term from spilled traffic.}
  \label{fig:cm_hw_arch}
    \vspace{-2mm}
\end{figure}

The traffic model (TM) captures how persistent decoder state translates into memory time. Message-passing decoders maintain state across iterations, including intermediate messages and auxiliary quantities \cite{roffe_2020,muller2025improvedbeliefpropagationsufficient}. By \emph{decoder state organization}, we refer to the choice of what information is stored across iterations. In this work, we compare two organizations. The first stores directed edge messages and therefore scales with the number of Tanner-graph edges $(E)$. The second stores cached summaries and beliefs and therefore scales with the number of variable and check nodes, that is, as $N+M$ up to constant factors and this difference is illustrated in Figure~\ref{fig:cm_hw_arch} block $a$. We denote the resulting persistent state footprint by $B_{\mathrm{state}}$. Unless explicitly swept, message and summary bit widths, storage-alignment granularity, and the bounded iteration budget are held fixed. Any footprint difference is therefore induced by state organization itself.

The available SRAM budget ($B_{\mathrm{SRAM}}$) determines whether the state fits on chip. We use a two-tier memory abstraction in which decoder state is served from on-chip SRAM up to $B_{\mathrm{SRAM}}$ or from a higher-latency off-chip tier once that budget is exceeded. On hardware platforms, this lower tier may correspond to external memory service when decoder storage exceeds available on-chip resources \cite{maurya2026fpgatailoredalgorithmsrealtimedecoding,das2021lilliputlightweightlowlatencylookuptable}. We define the excess state beyond the on-chip budget $(B_{\mathrm{excess}})$ as:
\begin{equation}
B_{\mathrm{excess}} \triangleq \max\{0,\; B_{\mathrm{state}} - B_{\mathrm{SRAM}}\}.
\label{eq:cm_excess_bytes}
\end{equation}
The SRAM fit boundary is the smallest budget for which $B_{\mathrm{excess}}=0$, as shown in block $b$ of Figure~\ref{fig:cm_hw_arch}. Because $B_{\mathrm{state}}$ includes alignment and granularity effects, the aligned persistent footprint determines the predicted fit-boundary location.

We model the off-chip traffic per iteration $(V_{\mathrm{off}}^{(\mathrm{iter})})$ and per job $(V_{\mathrm{off}}^{(\mathrm{tot})})$ as:
\begin{equation}
V_{\mathrm{off}}^{(\mathrm{iter})} \triangleq \rho_{\mathrm{rw}} \cdot B_{\mathrm{excess}},
\qquad
V_{\mathrm{off}}^{(\mathrm{tot})} \triangleq I \cdot V_{\mathrm{off}}^{(\mathrm{iter})},
\label{eq:cm_offchip_bytes}
\end{equation}
where $I$ is the bounded iteration budget per job and $\rho_{\mathrm{rw}}$ is an effective read/write amplification factor. As illustrated in Figure~\ref{fig:cm_hw_arch}, this first-order model captures the latency cost of serving excess state from off-chip memory rather than modeling a cycle-accurate memory system \cite{das2021lilliputlightweightlowlatencylookuptable}. Off-chip contention is not modeled explicitly as a separate memory queue; its aggregate effect is absorbed into the effective bandwidth $\beta$ (and any stall-related parameters), so it appears only through the memory-time term.

We convert traffic into a memory-time component using an effective bandwidth $\beta$ as:
\begin{equation}
S_{\mathrm{mem}} \triangleq \frac{V_{\mathrm{off}}^{(\mathrm{tot})}}{\beta}.
\label{eq:cm_mem_time}
\end{equation}
To prevent service time from vanishing in the fully on-chip regime, we include a small constant floor $S_{\mathrm{cmp},0}$ for fixed compute and dispatch overhead. The resulting traffic-model service time $(S_i^{(\mathrm{TM})})$ is defined as:
\begin{equation}
S_i^{(\mathrm{TM})} \triangleq S_{\mathrm{cmp},0} + S_{\mathrm{mem}}.
\label{eq:cm_tm_total}
\end{equation}
This decomposition appears in Figure~\ref{fig:cm_hw_arch} block $d$. As $B_{\mathrm{SRAM}}$ crosses the SRAM fit boundary, $B_{\mathrm{excess}}$ falls to zero, off-chip traffic collapses, and service time contracts under the same offered load. Bandwidth changes the scale of $S_{\mathrm{mem}}$ below the SRAM fit boundary without changing the fit-boundary location. The fit boundary is set by footprint relative to the on-chip budget.

\subsection{Workload Model}
\label{sec:cm_weight_v1}

The workload model (WM) captures compute-time variability across jobs. Let $w_i$ denote a scalar workload weight derived from the syndrome pattern $\in \{0, 1\}^{n-k}$ for an $[[n,k,d]]$ code in job $J_i$, such as the Hamming weight observed over the $W$-round window. Here, workload refers specifically to job-local decode difficulty summarized by $w_i$, not to the external arrival process or offered load. We model the corresponding compute-time contribution as:
\begin{equation}
S_{\mathrm{cmp},i}^{(\mathrm{WM})} \triangleq S_{\mathrm{base}} + \alpha w_i,
\label{eq:cm_wm_service}
\end{equation}
optionally capped to enforce a bounded per-job budget. WM is not intended to predict absolute decoder runtime. Instead, it introduces workload-correlated heterogeneity through a monotone proxy so that harder windows incur larger service demands. This creates controlled service-time variation and allows the evaluation to study latency tails induced by hard windows without introducing unbounded service.

\subsection{Composite Hardware Model}
\label{sec:cm_hw_v1}

The composite hardware model (CHM) combines memory and compute effects within a single service model. We define the compute component as:
\begin{equation}
S_{\mathrm{cmp}} \triangleq S_{\mathrm{base}} + \frac{U_i}{r_{\mathrm{cmp}}},
\label{eq:cm_chm_compute}
\end{equation}
where $U_i$ is a bounded work budget for job $J_i$ and $r_{\mathrm{cmp}}$ is the effective compute rate. The work budget $U_i$ may be fixed per job or derived from the workload weight $w_i$ of Sec.~\ref{sec:cm_weight_v1}.

We then combine compute and memory to obtain total service time, as shown schematically in block $d$ of Figure~\ref{fig:cm_hw_arch}:
\begin{equation}
S_i^{(\mathrm{CHM})} \triangleq
\begin{cases}
S_{\mathrm{cmp}} + S_{\mathrm{mem}}, & \textsc{Sum},\\
\max\{S_{\mathrm{cmp}}, S_{\mathrm{mem}}\}, & \textsc{Max}.
\end{cases}
\label{eq:cm_chm_combine}
\end{equation}
The \textsc{Sum} rule provides a conservative upper bound. The \textsc{Max} rule captures partial overlap when compute and memory can be pipelined. We use \textsc{Max} as the default and \textsc{Sum} as a pessimistic bound~\cite{openai_chatgpt_2026}.
\subsection{Reported Cost Quantities}
\label{sec:cm_metrics}

Alongside the service-level metrics of Sec.~\ref{subsec:pf_objectives_metrics}, we report cost quantities that explain misses, backlog growth, and latency tails in terms of memory and compute effects. For state footprint, we report the persistent state size $B_{\mathrm{state}}$, the excess state $B_{\mathrm{excess}}$ from Eq.~(\ref{eq:cm_excess_bytes}), and the fit-boundary location, $B^{\star} \triangleq \min\{B_{\mathrm{SRAM}} : B_{\mathrm{state}} \le B_{\mathrm{SRAM}}\}$, for each decoder organization. These quantities identify when the working state no longer fits on-chip. 

For traffic, we report off-chip bytes per iteration $V_{\mathrm{off}}^{(\mathrm{iter})}$ and total off-chip bytes per job $V_{\mathrm{off}}^{(\mathrm{tot})}$ from Eq.~(\ref{eq:cm_offchip_bytes}). We also report the corresponding traffic rate in bytes/s when characterizing load on the shared bottleneck. For service attribution, we report the memory and compute components, $S_{\mathrm{mem}}$ and $S_{\mathrm{cmp}}$, together with total service time under the selected model family. Time-based plots use total service time as the primary quantity, while component-wise times support attribution and diagnosis. These quantities help attribute observed misses and latency tails to off-chip memory cost induced by excess state or to bounded compute effort.
\subsection{Scope, Assumptions, and Limitations}
\label{sec:cm_assumptions}
The model families in this section are intended for comparative regime analysis rather than platform prediction. They isolate the effects of state footprint, excess state beyond the on-chip budget, workload heterogeneity, and compute--memory composition under shared assumptions while abstracting away platform-specific microarchitectural detail. Service-model constants should therefore be read as fixed operating points chosen to expose one system effect at a time, not as universal hardware claims. At the same time, they remain implementation-informed, reflecting bounded-iteration and reduced-precision decoding, microsecond-scale timing pressure, and decode-feedback paths for which backlog and reaction time matter \cite{caldwell2025platformarchitecturetightcoupling,maurer2025realtimedecodinggrosscode,bascones2025exploring,liu2026scalableopensourceqec}. The resulting conclusions are comparative and regime-oriented, identifying how decoder viability changes when memory fit, service variability, admission policy, or service capacity varies under otherwise controlled conditions. TM summarizes memory cost through effective bandwidth $\beta$ and an effective read/write amplification factor $\rho_{\mathrm{rw}}$. CHM captures compute--memory interaction through \textsc{Sum} and \textsc{Max} composition rather than through an explicit pipeline model.

The excess-state model uses a two-tier abstraction in which bandwidth affects memory time only after state exceeds the effective on-chip budget; the SRAM-fit boundary is determined by state footprint relative to that budget. WM uses a bounded scalar proxy for workload-dependent compute effort rather than an algorithm-specific runtime model. Reported service times should therefore be interpreted as comparative service-model quantities for regime analysis, not as platform-specific timing claims.


\begin{figure*}[tp]
    \centering
    \includegraphics[width=\textwidth, keepaspectratio]{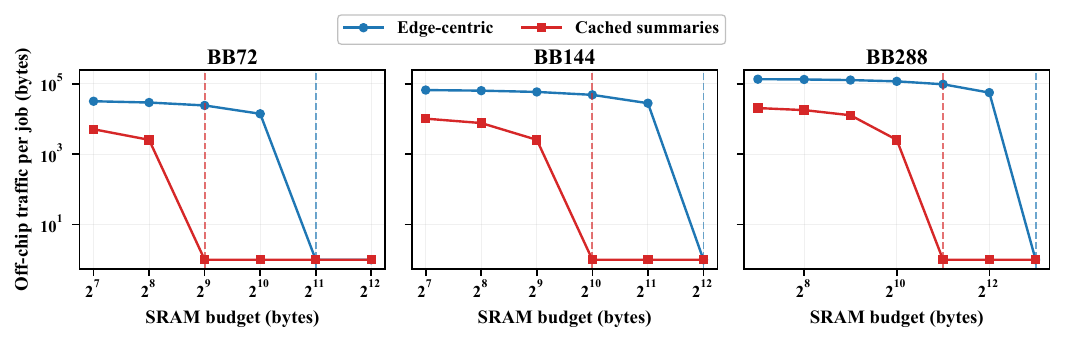}
    \caption{Off-chip traffic per decode job versus on-chip SRAM budget for edge-centric state
    and cached summaries across BB72, BB144, and BB288. Cached summaries shift the SRAM-Fit boundary left by \(4\times\) in all three workloads, causing off-chip traffic to collapse
    once the persistent decoder state fits on chip.}
    \label{fig:ev_claim1_knee}
    \vspace{-2mm}
\end{figure*}

\section{Experimental Setup}
\label{sec:methodology}


Using the cost models of Sec.~\ref{sec:cost_models_metrics}, we evaluate three controlled regimes together with a capacity-scaling extension. For each regime, we specify the service-model instantiation, operating point, sweep variables, and reproducibility controls while holding all other parameters fixed. The goal is to expose regime-level service behavior under fixed model assumptions rather than make platform-specific timing claims.

\subsection{Experimental Design}
\label{sec:em_design}
Our setup studies system drivers of deadline misses in real-time decoding together with the controls used to mitigate them. Across the study, the key factors are on-chip state budget ($B_{\mathrm{SRAM}}$), offered load, bounded decode-effort controls, and, when applicable, pooled decoder service capacity. We instantiate the framework using Belief Propagation-style decoder organizations \cite{roffe_2020,muller2025improvedbeliefpropagationsufficient,yao2024beliefpropagationdecodingquantum}, since their persistent iterative state provides a basis for studying state organization and SRAM-fit effects under parameterized service models. Decoder organization is varied only in the SRAM-fit-transition regime, where memory footprint is the variable of interest. In the remaining regimes it is held fixed to avoid confounding tail, overload, and capacity effects with footprint changes.

\subsubsection{SRAM-Fit-Transition Regime}
This regime is a cross-size memory-footprint study over the Bivariate Bicycle (BB) family \cite{Bravyi_2024} with $n \in \{72,144,288\}$ and $W=10$ extraction rounds under the traffic model of Sec.~\ref{sec:cm_traffic_v1}. It uses Poisson arrivals and slack fixed at \(50\,\mu\text{s}\). We compare two persistent-state organizations: an edge-centric layout that stores messages on graph edges, and a cached-summary layout that stores check summaries with variable-node beliefs. We sweep only $B_{\mathrm{SRAM}}$ while holding workload, deadlines, and service-model parameters fixed, isolating the shift in the SRAM-fit boundary and the transition in memory time.

\subsubsection{Tail-Latency Regime}
This regime uses BB72 with Poisson arrivals and slack fixed at \(100\,\mu\text{s}\) to study bounded cutoff and rescue policies under heavy-tailed service-time variability. Decoder organization is held fixed so that the observed tail behavior is not conflated with footprint changes.

\subsubsection{QoS Regime}
This regime uses BB144 with bursty on--off arrivals and slack fixed at \(100\,\mu\text{s}\) to study overload by sweeping the on-period arrival rate and backlog cap. This exposes the MissRate--DropRate tradeoff induced by admission control under a fixed service model.

\subsubsection{Capacity-Scaling Extension}
We keep the bursty BB144 QoS setting fixed and vary only the number of identical decoder service instances behind a shared queue. It isolates the effect of pooled decoder service capacity under overload. 

Together, these choices keep each regime compact while covering cross-size footprint effects, workload-sensitive tails, and overloaded queueing behavior.

\subsection{Service Models and Operating Points}
\label{sec:em_service_calibration}

This subsection specifies the cost-model instantiations used in each regime. The operating points are chosen to expose the target service-level transition under fixed assumptions. They are not cycle-accurate timing models for any particular decoder platform; instead, they provide bounded, comparable, implementation-informed settings for attributing effects to off-chip memory pressure, workload variability, admission control, and service capacity \cite{maurer2025realtimedecodinggrosscode,bascones2025exploring,liu2026scalableopensourceqec}.

\subsubsection{SRAM-Fit-Transition Regime}
For the SRAM-fit-transition study, we use the traffic model of Sec.~\ref{sec:cm_traffic_v1} with 10 BP iterations per admitted decode job, 16-bit quantization for messages, beliefs, and summaries, 64-byte state alignment, 64~GB/s effective memory bandwidth, and a \(0.5\,\mu\text{s}\) compute/dispatch floor. These settings define a bounded-iteration message-passing baseline with explicit alignment and bandwidth assumptions. The arrival process, deadline rule, and traffic-model parameters remain fixed. The comparison therefore attributes excess state beyond the on-chip budget, off-chip traffic, and fit-boundary movement to decoder state organization and on-chip budget.

\subsubsection{Tail-Latency Regime}
For the tail-latency study, we use the Composite Hardware Model (CHM) of Sec.~\ref{sec:cm_hw_v1}, with compute effort coupled to detector-event weight as in Sec.~\ref{sec:cm_weight_v1}. Total service time is composed using the \textsc{Max} rule of Eq.~(\ref{eq:cm_chm_combine}). The compute component includes a \(10\,\mu\text{s}\) baseline plus an additional term that scales with detector-event weight under fixed unit normalization. Heavy-tailed variability is introduced through Pareto service jitter with shape parameter \(2.0\), truncated at a \(50\times\) multiplier. These choices create a bounded but variable service regime in which cutoff and rescue selectivity can be evaluated. To isolate this effect, the embedded traffic baseline remains fully on-chip under the same 10-BP-iteration operating point. Bounded effort is then controlled through the cutoff budget and, when enabled, a fixed rescue budget.

\subsubsection{QoS Regime}
For the QoS-under-overload study, we suppress decoder-side variability by using a fixed per-job service time of \(20\,\mu\text{s}\). This keeps the service process constant, so the observed MissRate--DropRate tradeoff is attributable to bursty arrivals and backlog-cap admission control rather than service variability. The bounded-effort mechanisms used in the tail-latency study are disabled here.

\subsubsection{Capacity-Scaling Extension}
The capacity-scaling study reuses the same stressed bursty operating point as the QoS study. The per-instance service model remains fixed at \(20\,\mu\text{s}\) per job, and the workload, deadline rule, and arrival regime also remain fixed. Admission control, cutoff, and rescue are disabled, so the only changing factor is the number of identical decoder service instances behind the shared queue.

When model-derived timing components are reported, total service time remains the primary time quantity. Memory-only time is used only for attribution and diagnosis, while end-to-end latency metrics such as response time and its quantiles are defined in Sec.~\ref{subsec:pf_objectives_metrics}.

\subsection{Sweep Design and Reproducibility}
\label{sec:em_sweep_rigor}

Each regime sweeps only the variables needed to expose its target effect, while other regime parameters remain fixed.

\subsubsection{SRAM-Fit-Transition Regime}
In this study, the swept factor is the on-chip budget, $B_{\mathrm{SRAM}} \in \{128,256,512,1K,2K,4K,8K\}$~bytes. This range spans operating points from off-chip-memory-dominated to fully on-chip operation for the studied BB sizes.

\subsubsection{Tail-Latency Regime}
In this study, the heavy-tailed service regime is fixed while the cutoff budget is swept over $t \in \{20,30,50,70,100\}\,\mu\text{s}$. We compare four bounded-effort policies. Rescue-enabled runs use a fixed rescue budget of $10\,\mu\text{s}$, a backlog trigger of 2 jobs, and a slack trigger of $5\,\mu\text{s}$.

\subsubsection{QoS Regime}
In the QoS study, the burst structure and fixed-service regime remain constant while the on-period arrival rate is swept over $\lambda_{\mathrm{on}} \in \{20000,80000\}$ jobs/s and the backlog-cap threshold over $B_{\max} \in \{0,10,20,40,80,160,320\}$ jobs. These values span light to overloaded admission settings under the chosen burst process.

\subsubsection{Capacity-Scaling Extension}
In this study, we vary the number of identical decoder service instances with, $n_{\mathrm{servers}} \in \{1,2,4\}$, while the stressed bursty regime fixed.

Each reported sweep point corresponds to a fully specified configuration with recorded execution parameters and per-job outputs. For stochastic regimes, compared policies use the same seeded conditions. The SRAM-fit-transition, QoS, and capacity-scaling studies use 5000 jobs per run, whereas the tail-latency study uses 10000 jobs to stabilize tail summaries under heavier variability. Across the study, we report MissRate, DropRate, and \(p99\) response time, together with regime-specific quantities including off-chip traffic, trigger rate, goodput, and backlog statistics. Deterministic model-derived quantities, including predicted fit-boundary locations, are reported directly from the corresponding fixed model instantiations. Misses and drops are reported separately, consistent with Sec.~\ref{subsec:pf_objectives_metrics}, so overload control is not mistaken for slow service. All reported points are derived from per-job and queue-trajectory records logged by \framework.

\section{Evaluation}
\label{sec:evaluation}

\begin{figure*}[tp]
    \centering
    \includegraphics[width=\linewidth, keepaspectratio]{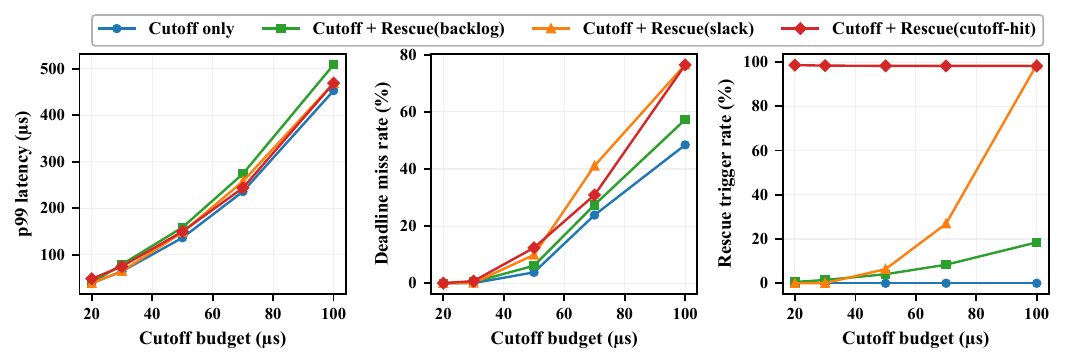}
    \caption{Tail-latency behavior under bounded mitigation policies. The three panels show p99
response time, MissRate, and rescue trigger rate versus cutoff budget for \emph{Cutoff only},
\emph{Cutoff + Rescue(backlog)}, \emph{Cutoff + Rescue(slack)}, and
\emph{Cutoff + Rescue(cutoff-hit)}.  The results show that bounded mitigation is policy-sensitive.
}
    \label{fig:ev_claim2_tail}
    \vspace{-2mm}
\end{figure*}


\subsection{SRAM Fit Boundary}
\label{subsec:ev_sram_knee}

We evaluate how decoder state organization changes the SRAM-fit boundary under the Traffic Model (TM) by sweeping the on-chip budget \(B_{\mathrm{SRAM}}\). Figure~\ref{fig:ev_claim1_knee} shows that cached summaries move the SRAM-fit boundary left by a consistent factor of \(4\times\) across BB72, BB144, and BB288: from 2048~B to 512~B, from 4096~B to 1024~B, and from 8192~B to 2048~B, respectively. This corresponds to a \(300\%\) increase in fully on-chip state capacity relative to the edge-centric layout. The sharp transition on the sweep grid shows that below the SRAM-fit boundary, off-chip traffic remains nonzero, whereas at and beyond the boundary it drops to zero. The observed boundary locations also match the aligned footprint accounting on the sweep grid, showing that the transition is explained by state size rather than by a secondary parameter interaction. This implies that, when the transition is footprint-driven, keeping the decoding latency low requires either reducing persistent decoder state through organization or increasing the available on-chip SRAM budget.

Once the persistent decoder state fits within \(B_{\mathrm{SRAM}}\), \(B_{\mathrm{excess}}\) becomes zero, so \(V_{\mathrm{off}}^{(\mathrm{iter})}\) and \(V_{\mathrm{off}}^{(\mathrm{tot})}\) collapse and the memory component of service time contracts accordingly. Cached summaries therefore create a wider fully on-chip operating region than the edge-centric layout. Crossing the fit boundary thus changes the service regime rather than producing a marginal traffic reduction, because it removes a dominant source of memory-induced service inflation and expands the feasible low-latency operating range.
\subsection{Tail-Latency Robustness}
\label{subsec:ev_tail_robustness}

\begin{table}[b]
    \centering
    \caption{Representative comparison of bounded mitigation policies at a cutoff budget of
    \(100~\mu\text{s}\).}
    \label{tab:ev_claim2_rep100}
    \scriptsize
    \renewcommand{\arraystretch}{1.1}
    \begin{tabular*}{\columnwidth}{@{\extracolsep{\fill}}lccc@{}}
        \hline
        Policy & MissRate & p99 latency & Trigger rate \\
        \hline
        Cutoff only & 48.44\% & 453~\(\mu\)s & 0.00\% \\
        Cutoff + Rescue(backlog) & 57.30\% & 509~\(\mu\)s & 18.49\% \\
        Cutoff + Rescue(slack) & 76.52\% & 469~\(\mu\)s & 98.83\% \\
        Cutoff + Rescue(cutoff-hit) & 76.44\% & 469~\(\mu\)s & 98.25\% \\
        \hline
    \end{tabular*}
\end{table}

The tail-latency regime evaluates bounded mitigation policies under heavy-tailed, workload-sensitive service variability at high load. Figure~\ref{fig:ev_claim2_tail} shows that mitigation helps only when the trigger remains selective. Across the sweep, \emph{Cutoff only} remains the strongest baseline, \emph{Cutoff + Rescue(backlog)} is the least harmful rescue policy, and \emph{Cutoff + Rescue(slack)} and \emph{Cutoff + Rescue(cutoff-hit)} degrade sharply as rescue becomes nearly always active. Under these latter triggers, rescue no longer acts as targeted recovery and instead behaves like persistent extra queue load.

Table~\ref{tab:ev_claim2_rep100} gives a representative comparison at a cutoff budget of \(100\mu\text{s}\). In the MissRate column, \emph{Cutoff only} performs best at 48.44\%, and \emph{Cutoff + Rescue(backlog)} is the closest rescue-based alternative at 57.30\%. By contrast, \emph{Cutoff + Rescue(slack)} and \emph{Cutoff + Rescue(cutoff-hit)} rise to 76.52\% and 76.44\%. The Trigger-rate column supports this interpretation: backlog-triggered rescue activates on only 18.49\% of jobs, whereas the slack- and cutoff-hit-triggered policies activate on 98.83\% and 98.25\% of jobs, or about \(5.3\times\) more often. The p99 latency column shows a different ordering: \emph{Cutoff + Rescue(backlog)} is highest at \(509~\mu\text{s}\), while the slack- and cutoff-hit-triggered policies are slightly lower at \(469~\mu\text{s}\) but substantially worse on MissRate. These results show that tail robustness is governed by rescue selectivity rather than by adding bounded auxiliary work. This implies that bounded auxiliary work is useful only when rescue remains selective; once trigger activity becomes persistent, tightening the trigger or disabling rescue is preferable to treating it as a generic tail-mitigation mechanism.

\begin{figure*}[t]
    \centering

    \begin{subfigure}[t]{0.24\textwidth}
        \centering
        \includegraphics[width=\linewidth]{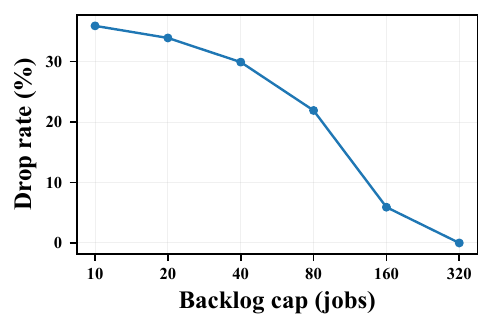}
        \caption{Drop rate under backlog-cap control ($\lambda_{\mathrm{on}}$=80000).}
        \label{fig:eval_claim3_drop}
    \end{subfigure}
    \hfill
    \begin{subfigure}[t]{0.24\textwidth}
        \centering
        \includegraphics[width=\linewidth]{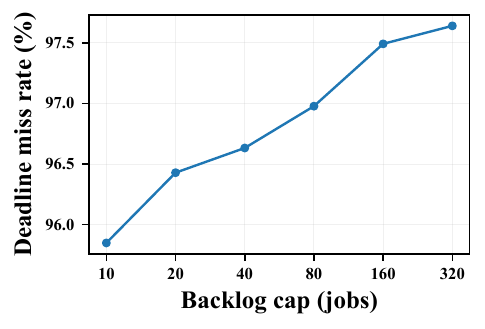}
        \caption{Deadline miss rate under backlog-cap control ($\lambda_{\mathrm{on}}$=80000).}
        \label{fig:eval_claim3_miss}
    \end{subfigure}
    \hfill
    \begin{subfigure}[t]{0.24\textwidth}
        \centering
        \includegraphics[width=\linewidth]{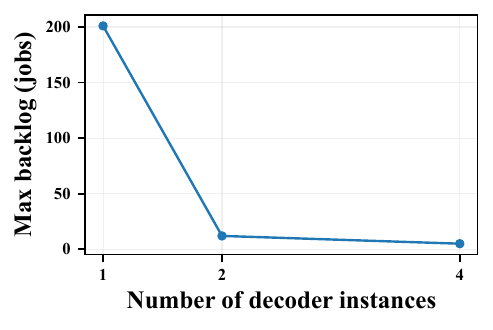}
        \caption{Maximum backlog vs number of decoder instances.}
        \label{fig:eval_claim4_backlog}
    \end{subfigure}
    \hfill
    \begin{subfigure}[t]{0.24\textwidth}
        \centering
        \includegraphics[width=\linewidth]{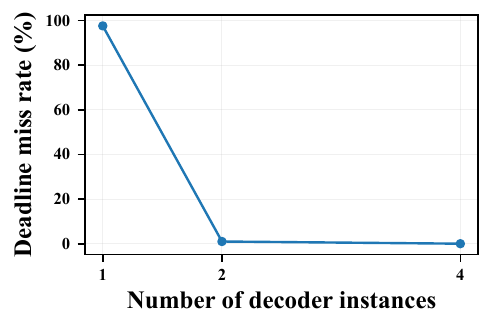}
        \caption{Deadline miss rate vs number of decoder instances.}
        \label{fig:eval_claim4_miss}
    \end{subfigure}

    \caption{QoS shaping and capacity scaling in the overloaded regime. 
    Top-level admission control trades drops against deadline misses by bounding queue growth, while increasing the number of decoder instances sharply reduces both backlog and misses.}
    \label{fig:eval_qos_and_pooling}
    \vspace{-2mm}
\end{figure*}
\subsection{QoS Under Overload}
\label{subsec:ev_qos_overload}

    


\begin{table}[b]
    \centering
    \caption{Representative operating points from the overloaded bursty QoS regime
    ($\lambda_{\mathrm{on}}=80000$).}
    \label{tab:ev_claim3_qos_points}
    \scriptsize
    \renewcommand{\arraystretch}{1.1}
    \begin{tabular*}{\columnwidth}{@{\extracolsep{\fill}}cccccc@{}}
        \hline
        Cap & DropRate & MissRate & Goodput & Max backlog & p99 latency \\
        \hline
        10  & 35.92\% & 95.85\% & 2.66\% & 10  & 220~\(\mu\)s \\
        80  & 21.92\% & 96.98\% & 2.36\% & 80  & 1.62~ms \\
        320 & 0.00\%  & 97.64\% & 2.36\% & 201 & 3.861~ms \\
        \hline
    \end{tabular*}
\end{table}

The QoS regime evaluates backlog-cap admission control under sustained bursty overload. Figures~\ref{fig:eval_claim3_drop} and~\ref{fig:eval_claim3_miss} show that relaxing the backlog cap reshapes overload but does not relieve it. As the cap increases, DropRate falls while MissRate rises, so the system moves from early rejection toward late failure. In this stressed regime, looser caps admit more work but do not recover timely service.

Table~\ref{tab:ev_claim3_qos_points} makes this tradeoff explicit at representative Cap settings. In the DropRate column, relaxing the cap from 10 to 320 lowers dropping from 35.92\% to 0.00\%. Over the same range, the MissRate column rises from 95.85\% to 97.64\%, so the additional admitted work does not produce deadline recovery. The Goodput column changes only from 2.66\% to 2.36\%, showing that useful throughput is nearly unchanged. The queueing cost is larger: the Max backlog column grows from 10 to 201 jobs, and the p99 latency column increases from \(220~\mu\text{s}\) to \(3.861~\text{ms}\), or about \(20.1\times\) more queued work and \(17.6\times\) worse tail latency. In this regime, admission control shapes overload rather than throughput. This implies that backlog caps should be chosen to set the preferred overload behavior, while nearly unchanged goodput across cap settings indicates that admission tuning alone cannot restore timely service. In the lower-rate sanity-check regime ($\lambda_{\mathrm{on}}=20000$), all cap settings behave nearly identically, confirming that the observed tradeoff is specific to true overload rather than a byproduct of the cap mechanism itself.

\subsection{Capacity Scaling}
\label{subsec:ev_capacity_scaling}


We evaluate pooled service capacity under the overloaded bursty setting. Figures~\ref{fig:eval_claim4_backlog} and~\ref{fig:eval_claim4_miss} show that the dominant transition occurs from one decoder instance to two. At that point, backlog growth collapses and deadline failure is nearly eliminated, indicating that the stressed operating point is capacity-limited. Increasing from two instances to four adds headroom rather than creating another regime change.

\begin{table}[b]
    \centering
    \caption{Representative operating points for pooling identical decoder instances in the
    overloaded bursty regime.}
    \label{tab:ev_claim4_pooling_points}
    \scriptsize
    \renewcommand{\arraystretch}{1.1}
    \begin{tabular*}{\columnwidth}{@{\extracolsep{\fill}}cccc@{}}
        \hline
        Instances & MissRate & Max backlog & p99 latency \\
        \hline
        1 & 97.64\% & 201 & 3.861~ms \\
        2 & 0.98\%  & 12  & 100~\(\mu\)s \\
        4 & 0.00\%  & 5   & 31~\(\mu\)s \\
        \hline
    \end{tabular*}
\end{table}
Table~\ref{tab:ev_claim4_pooling_points} makes this transition explicit with increasing instances in the Instances column. In the MissRate column, moving from one instance to two reduces the miss rate from 97.64\% to 0.98\%, and four instances eliminate misses in this experiment. The Max backlog column drops from 201 jobs to 12 and then to 5, showing that the first replication step removes most of the queued work. The p99 latency column falls from \(3.861~\text{ms}\) to \(100~\mu\text{s}\) and then to \(31~\mu\text{s}\), so the largest latency gain also occurs from one instance to two. Pooling therefore acts as an effective capacity lever in this overloaded setting. Operationally, pooling corresponds to replicating identical decoder engines or service slots behind a shared queue, so it complements rather than replaces per-decoder efficiency and robustness improvements. The trade-off is that improved timeliness is purchased by additional service capacity. This implies that, once a stressed operating point is identified as capacity-limited, recovering timely service requires additional decoder service capacity rather than further admission or rescue tuning. It also reinforces the need for queue-aware analysis beyond average-runtime decoder comparisons.


\label{subsec:ev_synthesis}



The evaluations show that real-time decoder viability is shaped by four factors: state organization, rescue semantics, admission control, and service capacity. Together, these results support modeling real-time QLDPC decoding as a deadline-driven online service rather than as an offline computation.


\section{Conclusion}
\label{sec:conclusion}


This work presented \framework{}, a deadline-driven systems framework for studying real-time QLDPC decoding under finite on-chip memory, service variability, overload, and limited service capacity. Across SRAM-fit-transition, tail-latency, overload, and capacity-scaling regimes, the results show that decoder viability is governed not only by correction quality or average runtime, but also by memory fit, rescue selectivity, admission policy, and service capacity. Cached summaries shift the fully on-chip SRAM-fit boundary left by \(4\times\). Bounded rescue remains effective only while its triggers stay selective. Under overload, relaxing the backlog cap causes about \(20.1\times\) more queued work and about \(17.6\times\) worse p99 latency with little gain in useful throughput. At capacity-limited operating points, pooling identical decoder instances reduces MissRate from 97.64\% to 0.98\% while sharply reducing backlog.

These findings support a systems view of real-time QLDPC decoding in which queueing, deadlines, and shared-resource limits are first-order design constraints. \framework{} therefore provides a controlled methodology for identifying the regime changes that determine whether a decoder remains viable in the control loop, with FPGA-based realization as a natural next step for hardware-grounded validation.

\section{Acknowledgment}
\label{sec:acknowledgment}

OpenAI ChatGPT~\cite{openai_chatgpt_2026} was used to assist with limited drafting and language organization in Sections~\ref{sec:problem_formulation}, \ref{sec:system_design}, and \ref{sec:cost_models_metrics}; all generated text was reviewed, revised, and verified by the authors.
\balance

\bibliographystyle{IEEEtran}
\bibliography{reference}

\end{document}